\documentclass[showpacs,amsmath,amssymb,twocolumn,superscriptaddress,notitlepage,preprintnumbers,pra]{revtex4-2}

\usepackage[dvips]{graphicx}
\usepackage{amsmath,amssymb,amsthm,mathrsfs,amsfonts,dsfont}

\usepackage{dcolumn}
\usepackage{bm}
\usepackage{amsmath}
\usepackage{amssymb}
\usepackage{braket}
\usepackage{physics}
\usepackage{amsthm}
\usepackage{natbib}
\usepackage{mathtools}
\usepackage{diagbox}
\usepackage{hyperref}
\usepackage[utf8]{inputenc}
\usepackage[english]{babel}
\usepackage{scalerel}[2014/03/10]
\usepackage{stackengine}
\usepackage{algorithm}
\usepackage{relsize}
\usepackage[noend]{algpseudocode} 

\usepackage{hyperref}
\hypersetup{colorlinks=true, linkcolor=blue, citecolor=blue, urlcolor=blue }

\usepackage{multirow}
\usepackage{makecell}
\usepackage{threeparttable}

\newcommand{\degree}{^\circ}

\usepackage[capitalize]{cleveref}

\newcommand{\SDU}{School of Physics, State Key Laboratory of Crystal Materials, Shandong University, Jinan 250100, China}

\begin{document}

\title{Chip-Integrated Broadband Multi-Photon Source for Wavelength-Multiplexed Quantum Networks}
\author{Xiao-Xu Fang}
\affiliation{\SDU}
\author{Ling-Xuan Kong}
\affiliation{\SDU}
\author{He Lu}
\email{luhe@sdu.edu.cn}
\affiliation{\SDU}


\begin{abstract} 
Quantum networks based on wavelength-multiplexed entanglement enable parallel distribution of quantum correlations, increasing channel capacity for secure communication and distributed quantum information processing. However, broadband integrated sources capable of generating multipartite entanglement beyond photon pairs remain scarce. Here we report on-chip generation of telecom-band four-photon entanglement in a periodically poled thin-film lithium niobate on insulator~(LNOI) waveguide. Type-0 spontaneous parametric down-conversion provides a phase-matching bandwidth exceeding 200~nm, enabling spectrally separable generation of multi-photon entanglement across the telecom band. The generated photons are encoded in time bins for robust fiber compatibility, and a coherent interface enabling reversible conversion between time-bin and polarization degrees of freedom allows complete quantum state tomography. We measure two-photon entanglement with a brightness of 6.7~MHz/mW/nm and a fidelity of $0.874 \pm 0.002$. At a pump power of 0.08~mW, the four-photon state exhibits a fourfold coincidence rate of 1~Hz and a fidelity of $0.74 \pm 0.01$, representing a threefold improvement over previous integrated platforms. Our results establish LNOI as a scalable platform for broadband multi-photon entanglement and provide a practical route toward dense wavelength-multiplexed quantum networks.
\end{abstract}

\maketitle

\noindent\textbf{\emph{Introduction.---}}The realization of large-scale quantum networks promises transformative capabilities in secure communication, distributed quantum computation, and quantum-enhanced sensing~\cite{kimble2008Nature,Simon2017NP,wehner2018Sa}. Early quantum communication experiments were primarily built on point-to-point links distributing two-photon maximally entangled states~\cite{Scarani2009RMP}. Recent advances in fully connected quantum network architectures enable wavelength-multiplexed entanglement distribution between arbitrary user pairs~\cite{Wengerowsky2018Nature,Siddarth2020SA,Alshowkan2021PRXQ,Appas2021npjQI,Wen2022PRApplied,Liu2022PX,Liu2024SA,Fan2025Light}. These developments establish a flexible and scalable framework for bipartite entanglement sharing across complex network topologies. 

As quantum networks continue to mature, extending their functionality beyond pairwise connectivity becomes a central objective. Enabling richer correlation structures among distant nodes requires the capability to generate and distribute multiple entangled pairs or genuine multipartite entangled states. Four-photon entanglement constitutes the minimal nontrivial resource for simultaneously establishing two Bell correlations or multipartite correlations across separate users. In particular, the simultaneous distribution of two Bell pairs underpins advanced protocols such as quantum network coding~\cite{Lu2019npjQI}, network nonlocality tests~\cite{Poderini2020NC,Carvacho2022Optica,Mao2023PRR,Meskine2025PRXQ}, and device-independent self-testing~\cite{AgrestiPRXQ2021}.

Realizing such multi-photon resources in practical fiber-based networks further demands an encoding scheme that preserves coherence over long distances while remaining compatible with scalable hardware. Time-bin encoding is especially attractive due to its intrinsic robustness against polarization fluctuations and phase noise in optical fibers~\cite{tittel1998PRLa,marcikic2003Na,marcikic2004PRLa,cuevas2013NCa,white2025PRLa}. By encoding quantum information in temporal modes, time-bin entanglement enables stable long-distance transmission without active polarization stabilization. Combining multi-photon entanglement with time-bin encoding therefore provides a resilient and scalable pathway toward advanced quantum network architectures.

Integrated photonic technologies provide a promising platform for the scalable generation of multi-photon entangled states, as demonstrated in high-index glass~\cite{reimer2016Sa} and silicon~\cite{zhang2019LSAa,Adcock2019NC,Feng2019npjQI,Llewellyn2020NP,Chen2023PRL,Lee2024APL}. Among various material platforms, thin-film lithium niobate on insulator~(LNOI) uniquely combines strong second-order nonlinearity with a compact and low-loss architecture~\cite{labbe2025APa}. Its intrinsic $\chi^{(2)}$ nonlinearity enables efficient spontaneous parametric down-conversion~(SPDC) within centimeter-scale devices~\cite{Fang2024LPR,Shi2024Light,Jiao2025PRL,chapman2025PRLa,shi2025arXiv,huang2025NP,wang2026arXiv}, while dispersion engineering supports broadband phase matching~\cite{Xue2021PRA, Javid2021PRL,Fang2024OE,fang2026OE}. These properties make LNOI particularly well suited for realizing broadband and power-efficient multi-photon sources for wavelength-multiplexed quantum networks.

\begin{figure*}[h!tbp]
\includegraphics[width=1\linewidth]{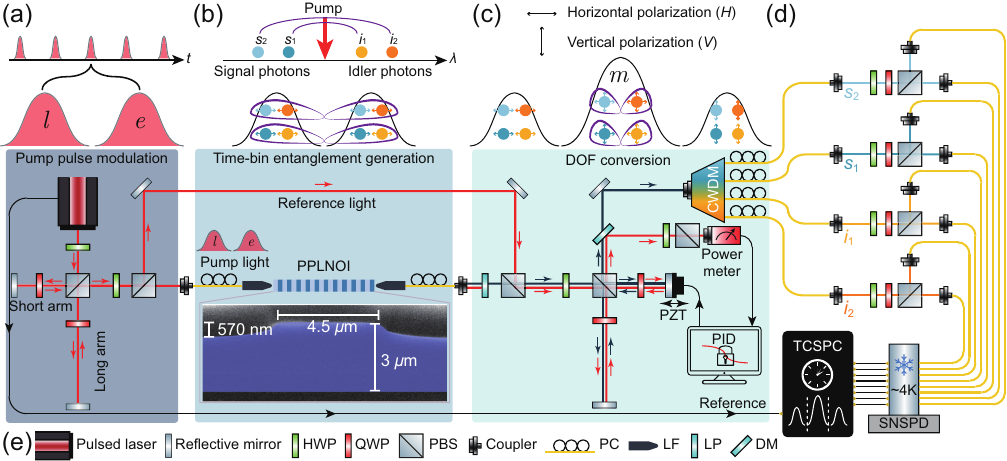}
\caption{Schematic of the experimental setup for the generation and characterization of multi-photon quantum states in a periodically poled lithium niobate (PPLN) waveguide. The experiment consists of four main stages: (a) modulation of the pump pulse train; (b) generation of time-bin–encoded multi-photon states in a shallow-etched PPLN waveguide; (c) coherent degree-of-freedom (DOF) conversion between time-bin and polarization encodings; and (d) polarization analysis and photon detection. (e) Symbols: HWP, half-wave plate; QWP, quarter-wave plate; PBS, polarizing beam splitter; PC, polarization controller; LF, lensed fiber; LP, long-pass filter; DM, dichroic mirror; TCSPC, time-correlated single-photon counting; SNSPD, superconducting nanowire single-photon detector.}
\label{Fig1:setup}
\end{figure*}

In this work, we demonstrate an efficient and broadband four-photon source on a thin-film LNOI platform. We design and fabricate a shallow-etched periodically poled lithium niobate~(PPLN) waveguide that enables high-efficiency photon-pair generation with a brightness of 6.7~MHz/mW/nm and a phase-matching bandwidth of 229~nm. Time-bin-encoded multi-photon states are generated using a temporally modulated pulsed pump. To achieve complete access to the generated quantum states, we develop a coherent degree-of-freedom (DOF) converter capable of transforming arbitrary states between time-bin and polarization encodings. Using this interface, we reconstruct the density matrices of both two- and four-photon entangled states, obtaining fidelities of $0.874 \pm 0.002$ and $0.74 \pm 0.01$, respectively. At a pump power of 0.08~mW, the four-photon state exhibits a fourfold coincidence rate of 1~Hz, representing an approximately threefold enhancement over previous integrated demonstrations.

\noindent\textbf{\emph{Experimental setup.---}}As illustrated in Fig.~\ref{Fig1:setup}, the experimental setup for generating and detecting the four-photon entangled state consists of three main parts. The first part prepares and modulates the pump light (Fig.~\ref{Fig1:setup}~(a)). A pulsed laser with a central wavelength of 775~nm, a repetition rate of 100~MHz, and a pulse duration of approximately 10 ps is used as the pump light. The pulses are injected into an unbalanced Mach–Zehnder interferometer~(UMZI), which introduces a fixed temporal delay of 645~ps between its two arms. Specifically, The horizontally polarized pump is first rotated by a half-wave plate~(HWP) set at 22.5$^\circ$, generating an equal superposition of horizontal and vertical polarizations. A polarizing beam splitter~(PBS) then separates the two polarization components, directing $\ket{V}$ into the short arm ($\ket{e}$) and $\ket{H}$ into the long arm ($\ket{l}$) of the UMZI. After propagation, the two components recombine at the PBS, yielding the time-bin-encoded pump state $\ket{\psi_p}=\frac{1}{\sqrt{2}}\left(\ket{e_p}\otimes\ket{H_p}+e^{i\phi_p}\ket{l_p}\otimes\ket{V_p}\right)$, where $\phi_p$ is determined by the optical path-length difference of the UMZI and is actively stabilized. The temporal separation (645~ps) is much larger than the pulse duration (10 ps), ensuring negligible temporal overlap between early and late time bins. At the same time, this delay remains within the coherence time of the pump laser, preserving phase coherence between the two components. 

After the UMZI, the pump passes through a second HWP (22.5$^\circ$) and a PBS. The reflected port serves as a reference beam for phase stabilization, while the transmitted component is prepared in the state $\ket{\psi_p}=\frac{1}{\sqrt{2}}\left(\ket{e_p}+e^{i\phi_p}\ket{l_p}\right)\otimes\ket{H_p}$. As shown in Fig.~\ref{Fig1:setup}~(b), the transmitted light is coupled into a single-mode fiber~(SMF), where its polarization is converted to vertical using a fiber polarization controller~(PC). The vertically polarized pump is then coupled into the PPLN waveguide via a lensed fiber to drive type-0 SPDC $\ket{V_p} \rightarrow \ket{V_{s_n}}\ket{V_{i_n}}$, where the subscripts $s_n$ and $i_n$ label the signal and idler photons of the $n$-th pair and satisfy energy conservation $\omega_p=\omega_{s_n}+\omega_{i_n}$. The PPLN waveguide is designed and fabricated on a 3-$\mu$m-thick $z$-cut 5\% MgO-doped LNOI. Shadow etching is adopted for broadband phase matching~\cite{Fang2024OE} (More details are provided in Appendix). Because the pump is prepared in a coherent superposition of early and late time bins, the SPDC process generates time-bin entangled photon pairs. For $n$ generated pairs, the state can be written as  
\begin{equation}\label{Eq:timebin}
\ket{\Phi_2}^{\otimes n}=\bigotimes_{n}\left[\frac{1}{\sqrt{2}}\left(\ket{e_{s_n}e_{i_n}}+e^{i\phi_p}\ket{l_{s_n}l_{i_n}}\right)\otimes\ket{V_{s_n}V_{i_n}}\right],
\end{equation}
where $\ket{e_{s_n}e_{i_n}}$ and $\ket{l_{s_n}l_{i_n}}$ denote the signal–idler photon pairs generated in the early and late time bins, respectively.

\begin{figure*}[ht!bp]
\includegraphics[width=1\linewidth]{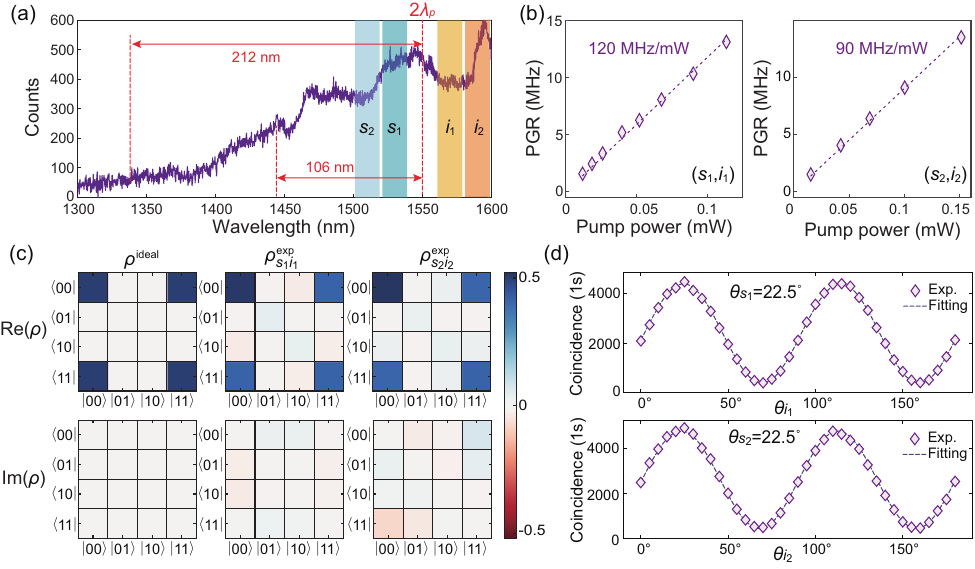}
 \caption{Characterization of the SPDC photon pairs. (a) Measured spectrum of the photons generated from SPDC.  
(b) Pair-generation rate~(PGR) at different pump powers for photon pairs $(s_1, i_1)$ and $(s_2, i_2)$.
(c) Real (Re) and imaginary (Im) parts of the theoretical and experimentally reconstructed density matrices of the four-photon entangled state for different spectral channels.  
(d) Two-photon interference fringes measured for photon pairs $(s_1, i_1)$ and $(s_2, i_2)$.}
\label{Fig2:photon2}
\end{figure*}

To characterize the generated time-bin entanglement, we convert it to polarization entanglement using a second UMZI (Fig.~\ref{Fig1:setup}~(c)).The signal and idler photons, together with the reference beam, are injected into this interferometer. Its path-length difference is precisely matched to that of the first UMZI, ensuring temporal overlap between the early–long and late–short time bins. Under this matched-delay condition, the time-bin entanglement described in Eq.~\ref{Eq:timebin} is coherently mapped onto the polarization DOF, resulting in
\begin{equation}
\ket{\Phi_2}^{\otimes n}=\bigotimes_n\left[\frac{1}{\sqrt{2}}\left(\ket{H_{s_n}H_{i_n}}+e^{i\phi}\ket{V_{s_n}V_{i_n}}\right)\otimes\ket{m_{s_n}m_{i_n}}\right], 
\end{equation} 
where $\phi=\phi_{s_n}+\phi_{i_n}-\phi_p$. The phases $\phi_{s_n}$ and $\phi_{i_n}$ arise from the optical path differences in the second UMZI. The state $\ket{m}$ corresponds to the middle time bin, formed by indistinguishable early–long and late–short paths. Photon pairs are postselected in this middle time bin using the pump pulse train as a timing reference, yielding a 50\% success probability for the time-bin-to-polarization conversion per channel. At the output of the second UMZI, the reference beam is separated from the photon pairs using a long-pass dichroic mirror filter. The reference light is detected by a power meter and fed into a proportional–integral–derivative (PID) feedback loop controlling a piezoelectric transducer (PZT) mounted on one interferometer mirror, thereby stabilizing the phase $\phi$. The generated photon pairs are separated using a coarse wavelength-division multiplexer~(CWDM). As shown in Fig.~\ref{Fig1:setup}~(d), the polarization states are analyzed using an HWP, quarter-wave plate~(QWP), and PBS. All photons are detected by superconducting~nanowire single-photon detectors~(SNSPDs), and coincidence events are recorded using a time-correlated single-photon counting~(TCSPC) system.

\begin{figure*}[ht!bp]
\includegraphics[width=1\linewidth]{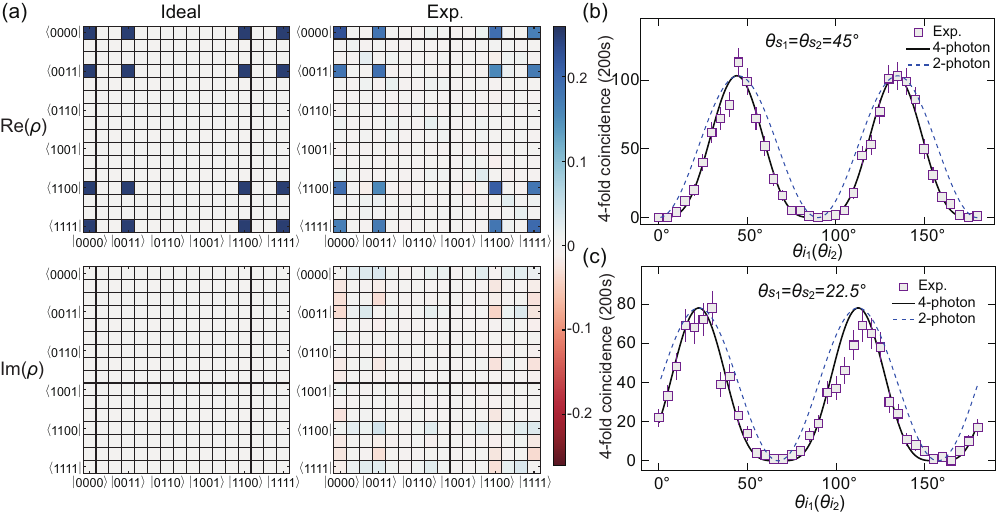}
\caption{Characterization of the four-photon quantum state. (a) Real (Re) and imaginary (Im) parts of the theoretical and experimentally reconstructed density matrices of the four-photon entangled state.
(b) Four-photon interference measured in the $\ket{H}/\ket{V}$ basis and the $\ket{\pm}$ basis.  The solid curve represents the theoretical four-photon interference prediction, while the dashed curve shows the cosine dependence expected for two-photon interference.}
\label{Fig3:photon4}
\end{figure*}

\noindent\textbf{\emph{Entanglement characterization.---}}We first characterize the broadband SPDC emission from the PPLN waveguide. As shown in Fig.~\ref{Fig2:photon2}~(a), the source exhibits a smooth and continuous spectrum with a 3-dB bandwidth of 229 nm and a base-to-base bandwidth of 504 nm. This exceptionally broad phase-matching window enables simultaneous access to multiple coarse wavelength-division multiplexing~(CWDM) channels, supporting parallel entanglement distribution across the telecom band. Within this bandwidth, we select two symmetric signal–idler channel pairs with respect to 2$\lambda_p$, denoted as $(s_1, i_1)$ and $(s_2, i_2)$, enabling parallel generation of spectrally separated entangled photon pairs. The correlated photons are separated using CWDM filters centered at 1510~nm, 1530~nm, 1570~nm, and 1590~nm, each with an 18-nm 3-dB bandwidth. The measured joint spectral intensities of both channel pairs exhibit well-defined anti-diagonal structures, confirming strong energy conservation in the SPDC process (see Appendix). For each channel pair, we measure the signal, idler and two-fold coincidence rate at different on-chip pump power, and extract the photon-pair generation rate~(PGR)~(see Appendix for details). As shown in Fig.~\ref{Fig2:photon2}~(b), the PGRs are 120 MHz/mW for $(s_1, i_1)$ and 90 MHz/mW for $(s_2,i_2)$. Normalizing by the 18-nm CWDM bandwidth yields spectral brightness values of 6.7 MHz/mW/nm for $(s_1, i_1)$ and 5 MHz/mW/nm for $(s_2,i_2)$. Given the 100~MHz repetition rate of the pump laser, the average photon-pair generation rate per pulse is estimated to be on the order of $10^{-1}$ photon pairs at a pump power of 0.1~mW. 

At a pump power of 0.12~mW, we reconstruct the experimental density matrices $\rho^\text{exp}_{s_1i_1}$ and $\rho^\text{exp}_{s_2i_2}$ via quantum state tomography\cite{Kwiat2005AP}, as shown in Fig.~\ref{Fig2:photon2}~(c). The fidelities with respect to the ideal Bell state $\ket{\Phi_2}$ are calculated as $\mathcal F_{s_ni_n}=\Tr\left(\rho^\text{expt}_{s_ni_n}\ket{\Phi_2}\bra{\Phi_2}\right)$, yielding $\mathcal F_{s_1i_1}=0.874 \pm 0.002$ and $\mathcal F_{s_2i_2}=0.872 \pm 0.002$. We further verify entanglement via polarization-correlation measurements. The angles of HWPs in the signal photons are fixed at $\theta_{s_1},\theta_{s_2}=22.5^\circ$, while the angles of HWPs in the idler photons are rotated from 0 to 180$^\circ$. The measured two-fold coincidence counts, shown in Fig.~\ref{Fig2:photon2}~(d), exhibit clear sinusoidal oscillations. The visibility is calculated by $\mathcal V=(C_\text{max}-C_\text{min})/(C_\text{max} + C_\text{min})$, where $C_{\max}$ and $C_{\min}$ are obtained from sinusoidal fits to the data. The visibility is  84\% for $(s_1, i_1)$ and 83\% for $(s_2, i_2)$. Both values exceed the threshold $1/\sqrt{2} \approx 0.71$, confirming entanglement through violation of the Clauser–Horne–Shimony–Holt inequality~\cite{Clauser1969PRLCHSH}.

Having thoroughly characterized the two-photon entanglement, we next investigate the four-photon state arising from the simultaneous generation of two photon pairs within a single pump pulse. At a pump power of approximately 0.08~mW, we observe a four-fold coincidence rate of 1~Hz. The reconstructed density matrix $\rho^\text{expt}$ is shown in Fig.~\ref{Fig3:photon4}~(a). The fidelity with respect to ideal state $\ket{\Phi_4}=\ket{\Phi_2}^{\otimes2}$ is calculated as $\mathcal F=\Tr\left(\rho^\text{expt}\ket{\Phi_4}\bra{\Phi_4}\right)=0.74\pm0.01$. To further verify multipartite coherence, we measure the four-photon polarization correlations. In Fig.~\ref{Fig3:photon4}~(b), the angles of HWPs in the signal photons are fixed at $\theta_{s_1}=\theta_{s_2}=45^{\circ}$, corresponding to projection in the $\ket{H}/\ket{V}$ basis, while the angles of HWPs in the idler channels $i_1$ and $i_2$ are rotated synchronously from $0^{\circ}$ to $180^{\circ}$. The recorded four-fold coincidences exhibit high-contrast interference fringes, yielding a visibility of 99.9\%. 
In Fig.~\ref{Fig3:photon4} (c), the angles of HWPs in the signal photons are fixed at $\theta_{s_1}=\theta_{s_2}=22.5^{\circ}$, corresponding to projection in the $\ket{\pm}=\frac{1}{\sqrt{2}}(\ket{H}\pm\ket{V})$ basis, while the idler HWPs are again rotated synchronously. The resulting four-photon interference yields a visibility of 95.9\%. The observed interference patterns agree well with the theoretical prediction for four-photon entangled state~(black solid lines) and are clearly distinct from two-photon interference fringes~(blue dashed lines). The consistently high visibilities in complementary measurement bases confirm the nonclassical correlations of the generated four-photon state.

\noindent\textbf{\emph{Conclusion and discussion.---}}
In conclusion, we experimentally demonstrate broadband four-photon entanglement on a thin-film LNOI platform. By leveraging type-0 SPDC and time-bin encoding, we generate telecom-band four-photon entangled states with a fidelity of $0.74 \pm 0.01$ and a four-fold coincidence rate of 1~Hz, representing an approximately threefold enhancement compared with previous implementations. The measured phase-matching bandwidth exceeding 200~nm provides intrinsic compatibility with dense wavelength-division multiplexing, enabling parallel entanglement distribution across multiple frequency channels. These results establish LNOI as a scalable and practical $\chi^{(2)}$ platform for integrated multi-photon quantum photonics and constitute an important step toward chip-based wavelength-multiplexed quantum networks and large-scale entanglement distribution.

Compared with previously reported four-photon entanglement sources based on other material platforms, our results enhance the four-fold coincidence rate by approximately a factor of three (see Appendix for detailed comparison). In the present experiment, the total loss per channel is approximately 15.5 dB, with a detailed breakdown provided in the Appendix. The fiber-to-chip coupling loss contributes about 4~dB, which can be reduced to below 0.5 dB using state-of-the-art taper designs and optimized fabrication processes \cite{Liu2022APN,Kashiwazaki2021APL}. In addition, the 3 dB loss associated with DoF conversion originates from probabilistic time-bin post-selection and is not fundamental. This loss can be eliminated by implementing deterministic time-bin analyzers based on electro-optic modulation~\cite{Vedovato2018PRL, Santagiustina2024Optica, bacchi2025arxiv}, which are fully compatible with the LNOI platform. With these improvements, the fourfold coincidence rate is expected to increase by up to two orders of magnitude, substantially enhancing multi-photon generation efficiency.



\appendix
\onecolumngrid

\section{Design and fabrication of the \\ periodically poled lithium niobate~(PPLN) waveguide}
\begin{figure*}[htb]
\includegraphics[width=1\linewidth]{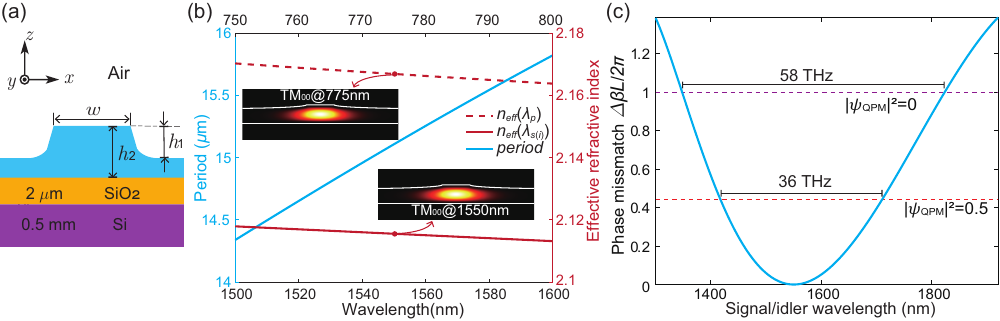}
\caption{Numerical simulation. (a) The cross section of the designed waveguide. (b) Simulated results. Both the near-infrared pump and telecom-band signal (idler) operate in the fundamental quasi-transverse-magnetic mode (TM$_{00}$), with the corresponding effective refractive indices $n_{\text{eff}}(\lambda_p)$ (red dashed line) and $n_{\text{eff}}(\lambda_{s(i)})$ (red solid line) plotted as functions of wavelength. The blue line represents the calculated QPM period $\Lambda$. The insets are the mode profiles of pump light at 775 nm and signal (idler) light at 1550 nm. (c) The phase mismatching $\Delta \beta L/2\pi$ is plotted with blue solid line for a waveguide length $L=12~mm$ with an appropriate choice of poling period. Purple dashed line indicates the first zero of the phase-mismatch sinc function and red dashed line indicates the $|\psi_\mathrm{QPM}|^2=0.5$.}
\label{FigS1:chipdesign}
\end{figure*}

We consider spontaneous parametric down-conversion (SPDC) in a straight ridge waveguide fabricated on an lithium-niobate-on-insulator~(LNOI) platform consisting of a 3-$\mu$m-thick $z$-cut 5\% MgO-doped lithium niobate (LN) thin film bonded to a 0.5-mm-thick silicon substrate via a 2-$\mu$m-thick SiO$_2$ layer. The simulated waveguide cross section is shown in Fig.~\ref{FigS1:chipdesign} (a), with a top width of $w = 4.5~\mu$m, a thickness of LN $h_2 = 3~\mu$m and an etch depth of $h_1 = 0.58~\mu$m. This geometry is chosen to support fundamental transverse-magnetic (TM$_{00}$) modes at both the pump and down-converted wavelengths. In the degenerate SPDC process, a pump photon ($p$) at angular frequency $\omega_p$ is converted into two photons—signal ($s$) and idler ($i$)—with equal frequencies $\omega_s = \omega_i = \omega_p/2$. Efficient nonlinear interaction requires satisfaction of the quasi-phase-matching (QPM) condition $\Delta \beta = 2\beta_{s(i)} - \beta_p -\beta_{\Lambda}  = 0$, where $\beta = \frac{2\pi}{\lambda} n_{\mathrm{eff}}(\lambda)$ is the propagation constant, $n_{\mathrm{eff}}$ is the effective refractive index, and $\beta_{\Lambda} = \frac{2\pi}{\Lambda}$ is the reciprocal lattice vector introduced by periodic poling with period $\Lambda$. To determine the poling period required for the type-0 phase matching ($\text{TM}_{00}\to\text{TM}_{00}+\text{TM}_{00}$), we numerically calculate the effective refractive indices of the fundamental TM$_{00}$ modes at the pump wavelength $\lambda_p$ and the degenerate signal/idler wavelength $\lambda_{s(i)} = 2\lambda_p$. The simulated effective indices for $n_{\text{eff}}(\lambda_p)$ in the range 750–800 nm are shown as a red dashed curve in Fig.~\ref{FigS1:chipdesign} (b), while the corresponding $n_{\text{eff}}(\lambda_{s(i)})$ values are shown as a red solid curve. For $\lambda_p = 775$ nm and $\lambda_{s(i)} = 1550$ nm, the calculated effective refractive indices are $n_\text{eff}(775nm) \approx 2.1669$ and $n_\text{eff}(1550nm) \approx 2.1152$. 
The simulated TM$_{00}$ mode profiles at both wavelengths are shown in the insets of Fig.~\ref{FigS1:chipdesign} (b), indicating strong spatial mode confinement and overlap between pump and down-converted fields. 
Accordingly, we can calculate the QPM period
\begin{equation}
    \Lambda = \frac{775~nm}{n_\text{eff}(775nm)-n_\text{eff}(1550nm)}\approx 15 \mu\text{m}. 
\end{equation}
By setting the waveguide length $L = 12$~mm, we calculate the phase mismatching $\Delta \beta L/2\pi$ as shown in Fig.~\ref{FigS1:chipdesign} (c). When $\Delta \beta L/2\pi=1$, the QPM function $|\psi_\mathrm{QPM}|^2=|\mathrm{sinc}(\Delta \beta L/2)|^2=0$ (marked with a purple dashed line)
indicates the largest phase-matching bandwidth of 58~THz. Considering the non-uniformity in the fabricated waveguide structure, the actual phase-matching bandwidth increases, though this comes with a decrease in brightness. Besides, the full width at half maximum (FWHM) of the phase-matching function is approximately 36~THz (marked with a red dashed line).

\begin{figure*}[htbp]
\includegraphics[width=0.8\linewidth]{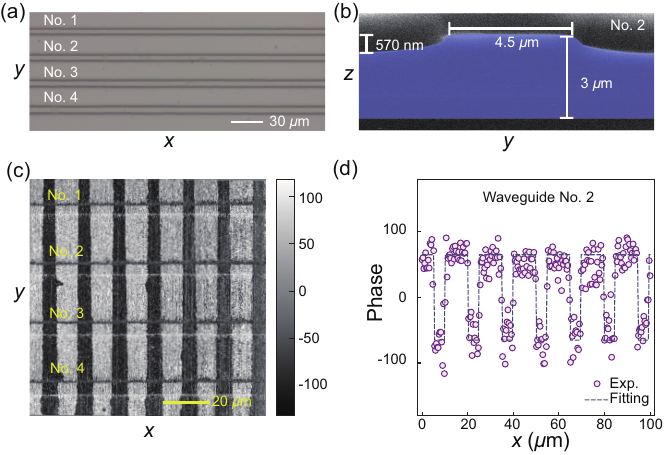}
\caption{Chip fabrication. (a) Top view of the fabricated chip under an optical microscope. (b) The SEM images of the end view of the experimental waveguide. (c) The PFM images of the poling region. (d) Duty cycle of domain inversion along the experimental waveguide No.~2.}
\label{FigS2:chip}
\end{figure*}

To fabricate the PPLN waveguides, we first define the electrode patterns by laser direct writing, followed by deposition of an 80-nm-thick Cr layer and lift-off to form the poling electrodes. Each electrode finger has a length of 125 $\mu$m, providing a sufficiently large poled region to accommodate multiple waveguide structures. During the poling process, the patterned Cr electrodes on the LN surface serve as the anode, while the low-resistivity silicon substrate functions as the ground electrode. Voltage pulses of approximately 800 V with a duration of 200 ms per pulse are applied to induce domain inversion along the $z$ axis, resulting in a periodically poled LNOI sample. After poling, the Cr electrodes are removed using a metal etchant.
Waveguide fabrication is subsequently carried out within the poled regions. The waveguide patterns are defined by laser direct writing, followed by deposition of a 200-nm-thick Cr hard mask. The pattern is transferred into the LN thin film via inductively coupled plasma reactive ion etching (ICP-RIE) to form ridge waveguides. During ICP-RIE, the ICP power is set to 700 W and the RIE power to 100 W. A gas mixture of SF$_6$ and Ar with a ratio of 3:1 is employed. The etch rate is approximately 8.5 nm/min, with an etch selectivity exceeding 3:1 relative to the Cr mask.
Four waveguides are fabricated within the poled region, as illustrated in Fig.~\ref{FigS2:chip}~(a). In our experiment, waveguide No. 2 is selected for entanglement generation. A scanning electron microscope (SEM) image of the end facet of waveguide No. 2 is shown in Fig.~\ref{FigS2:chip}~(b), revealing a well-defined ridge structure. The inverted ferroelectric domains are characterized by piezoresponse force microscopy (PFM), as shown in Fig.~\ref{FigS2:chip}~(c), where the inverted domains appear in light gray. The PFM data along waveguide No. 2 are presented in Fig.~\ref{FigS2:chip}~(d). From square-wave fitting of the domain pattern, the inverted-domain duty cycle is estimated to be approximately 0.68.

\section{Characterization of the PPLN waveguide}
\subsection{Second harmonic generation}

\begin{figure}[htbp]
\includegraphics[width=0.5\linewidth]{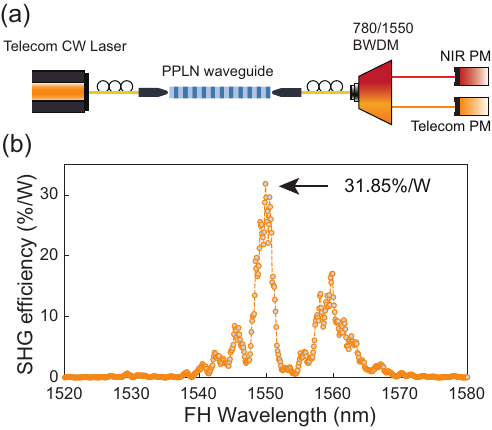}
\caption{Second harmonic generation~(SHG) characterization of the PPLN waveguide. (a) Experimental setup for measuring the SHG efficiency of the waveguide. 780/1550 BWDM, 780/1550~nm band wavelength division multiplexer; PM, power meters. (b) The measured SHG efficiency varying with the FH wavelength. The peak conversion efficiency is approximately 31.85\%/W.}
\label{FigS3:shg}
\end{figure}

Prior to SPDC characterization, the classical nonlinear conversion efficiency of the PPLN waveguide is evaluated by second-harmonic generation (SHG). As shown in Fig.~\ref{FigS3:shg} (a), the first-harmonic~(FH) light in the telecom band is provided by a continuous-wave~(CW) tunable laser~(Santec, TSL550). A polarization controller~(PC) is used to ensure the FH light maintained TM polarization, which is then coupled into the PPLN waveguide via a lensed fiber~(LF). The generated second-harmonic~(SH) light, along with the residual FH light, is coupled out through another LF and subsequently separated by a 775/1550~nm band wavelength division multiplexer~(BWDM). The power intensities of FH and SH lights~(denoted as $P_\text{FH}$ and $P_\text{SH}$, respectively), are recorded with a telecom power meter~(PM) and a near infrared~(NIR) PM, respectively. The SHG efficiency $\eta_\text{SHG}$ is calculated by
\begin{equation}
\eta_{\text{SHG}}=\frac{P_\text{SH}/\eta_\text{SH}}{(P_\text{FH}/\eta_\text{FH})^2},
\end{equation}
where $\eta_\text{FH}$ and $\eta_\text{SH}$ are the transmission efficiencies for FH and SH lights, respectively. With the PPLN chip is mounted on a thermometric cooler set at $21^{\circ}$C, we obtain the SHG spectrum~(in terms of $\eta_{\text{SHG}}$) by sweeping the wavelength of the FH light as shown in Fig.~\ref{FigS3:shg}~(b). The peak conversion efficiency is approximately 31.85\%/W.

We calculate the peak theoretical value the designed PPLN waveguide by 
\begin{align}
\eta_{\text{SHG}}^\text{th}=\frac{32 d_{33}^2 L^2}{\varepsilon_0 c n_{\text{FH}}^2 n_{\text{SH}} \lambda_{\text{FH}}^2} \frac{\zeta^2}{A_{\mathrm{eff}}}\left|{\text{sin}(\pi D)}\right|^2,
\end{align}
where nonlinear coefficient $d_{33}$=27~pm/V, polarization domain length $L = 12~mm$, $\varepsilon_0$ is the vacuum permittivity and $c$ is the speed of light in vacuum. For the wavelength of FH light $\lambda_{\text{FH}}=1550$nm, the effective refractive indices $n_{\text{FH}}=n_\text{eff}(1550nm) \approx 2.1152$, $n_{\text{SH}}=n_\text{eff}(775nm) \approx 2.1669$ . The effective model area is $A_{\text {eff}}=\left(A_\text{FH}^2 A_{\text{SH}}\right)^{\frac{1}{3}}$ and $A_{\text{FH}}$ and $A_{\text{SH}}$ are the effective model profiles for TM mode at 1550~nm and 775~nm, respectively. $\zeta$ represents the spatial mode overlap factor between the two modes. Considering the duty cycle is $D=0.68$, we calculate $\eta_{\text{SHG}}^\text{th}=260\%/W$.
The discrepancy between the experimental and simulation results is primarily attributed to imperfect waveguide structure and propagation loss.

\subsection{SPDC}

\begin{figure*}[htb]
\includegraphics[width=1\linewidth]{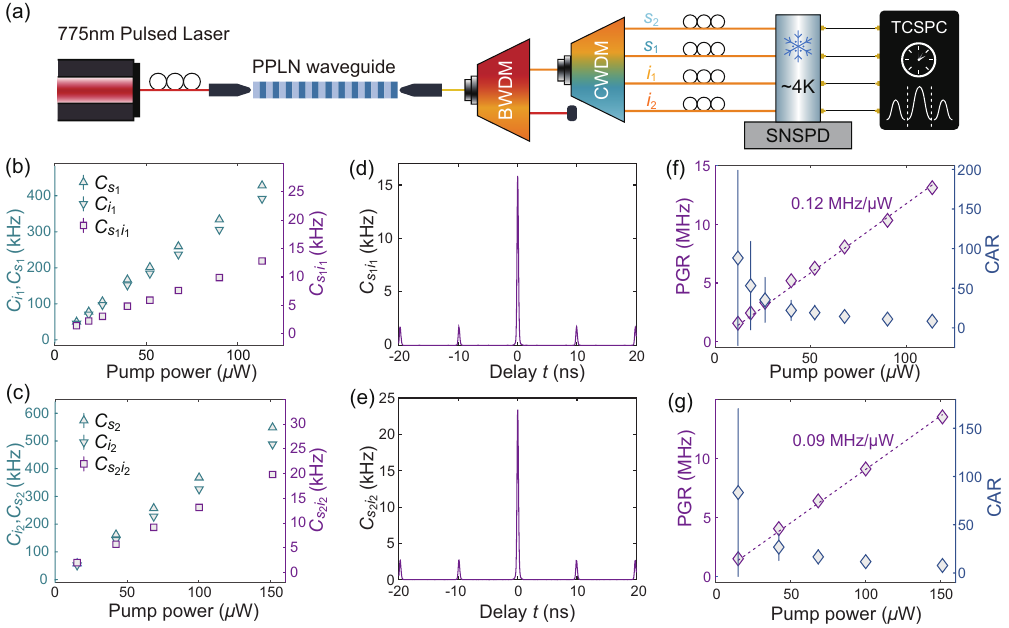}
\caption{Photon-pair generation~(PGR) and coincidence-to-accidental ratio~(CAR) characterization of photon pairs. (a) Experimental setup. A 775~nm pulsed laser directly pumps the PPLN waveguide, generating two photon pairs $(s_1, i_1)$ and $(s_2, i_2)$ used for our experiment. A BWDM is used to filter out the residual pump light. The generated photons are demultiplexed into four independent channels by a coarse wavelength-division multiplexer~(CWDM) and then detected by superconducting nanowire single-photon detectors~(SNSPDs). A time-correlated single-photon counting~(TCSPC) records two-fold coincidences for each of the two photon pairs separately. (b) and (c) Signal (Idler) count rate $C_{s_n}$ ($C_{i_n}$) and coincidence count rate $C_{s_n i_n}$ from $n$-th photon pairs with various pump power. (d) and (e) The signal-idler coincidence counts with different time delays $t$ of photon pairs $(s_1, i_1)$  at the pump power of 113~$\mu$W and photon pairs $(s_2, i_2)$ at the pump power of 151~$\mu$W, respectively. (f) and (g) The PGR and CAR values with various pump power of photon pairs $(s_1, i_1)$ and $(s_2, i_2)$, respectively. The error bars are one standard deviation estimated by including the Poisson statistics of the emitted photons and the fluctuations. }
\label{Fig4:PGR}
\end{figure*}

We use the experimental setup shown in Fig.~\ref{Fig4:PGR} (a) to characterize the photon-pair generation~(PGR) rate and coincidence-to-accidental ratio~(CAR) properties of the PPLN waveguide. A 775nm pulsed laser directly pumps the waveguide. The two photon pairs $(s_1, i_1)$ and $(s_2, i_2)$ used for our experiment are detected after BWDM filtering and coarse wavelength-division multiplexer~(CWDM) demultiplexing. Data is acquired at various pump powers, exhibiting an approximately linear scaling with the pump power, as shown in Fig.~\ref{Fig4:PGR} (b) and (c). We denote the count rates of signal and idler photons from the $n$-th pair as $C_{s_n}$ and $C_{i_n}$, respectively, and the coincidence count rate as $C_{s_n i_n}$. The PGR is then calculated by 
\begin{equation}
    \text{PGR} = \frac{C_{s_n} C_{i_n}}{C_{s_n i_n}}. 
\end{equation}
By fitting the slope in Fig.~\ref{Fig4:PGR} (f) and (g), we obtained an on-chip PGR of 0.12~MHz/$\mu$W for $(s_1, i_1)$ and 0.09~MHz/$\mu$W for $(s_2, i_2)$, respectively. Dividing the PGR by the 3-dB bandwidth of the CWDM (18~nm) gives the spectral brightness values of 6.7~MHz/mW/nm for $(s_1, i_1)$ and 5~MHz/mW/nm for $(s_2, i_2)$. 

The temporal correlation of signal and idler photons is characterized by the CAR, which is obtained by measuring $C_{s_n i_n}$ by introducing a time delay $t$ between signal and idler photons of each pair. Fig.~\ref{Fig4:PGR} (d) and (e) show $C_{s_n i_n}(t)$ at the pump power of 113~$\mu$W for $(s_1, i_1)$ and 151~$\mu$W for $(s_2, i_2)$, respectively. The CAR is then calculated by
\begin{equation}
    \text{CAR} = \frac{C_{s_n i_n}(0) - C_{s_n i_n}(\Delta t)}{C_{s_n i_n}(\Delta t)}, 
\end{equation}
where $C_{s_n i_n}(0)$ denotes the coincidence count rate for signal and idler photons generated within the same pump pulse ($n=0$, zero time delay), and $C_{s_n i_n}(\Delta t)$ is the accidental background rate measured at time delays of $\Delta t = n T$ (where $n = \pm1, \pm2, \ldots$ and $T = 1 / f_{\text{rep}}$ is the laser pulse period). Experimentally, we use $\Delta t = \pm 10$~ns, corresponding to $n = \pm1$ for $f_{\text{rep}} = 100$~MHz, where temporal photon correlations vanish and only accidental coincidences are detected. The CAR value recorded as a function of pump power is depicted in Fig.~\ref{Fig4:PGR} (f) and (g), which decreases with increasing PGR due to higher-order excitations in SPDC, which correspondingly reduces the temporal correlation.

\begin{figure*}[htb]
\includegraphics[width=1\linewidth]{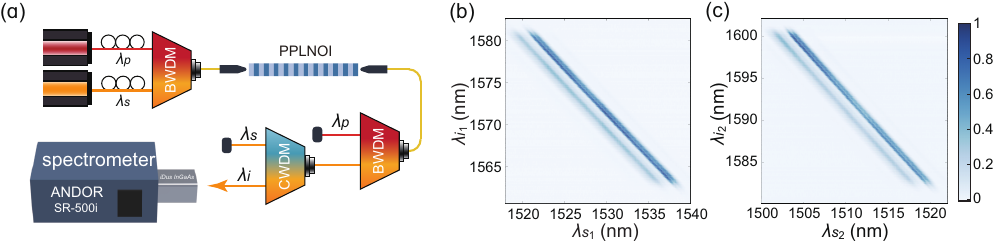}
\caption{Joint spectral intensity~(JSI) characterization of photon pairs. (a) Experimental setup. A 775~nm pulsed laser ($\lambda_p$) and a telecom CW laser ($\lambda_s$) are combined by a BWDM, with $\lambda_s$ being scanned within the wavelength range of the CWDM channel corresponding to the signal photon. The PPLN waveguide is simultaneously pumped by the two wavelengths, generating idler photons with the wavelength of $\lambda_i$ through DFG. After filtering out $\lambda_p$ and $\lambda_s$ using a BWDM and CWDM, $\lambda_i$ is directed into a spectrometer to measure its spectral information. (b) and (c) JSI of photon pairs $(s_1, i_1)$ and $(s_2, i_2)$, respectively. }
\label{FigS5:JSI}
\end{figure*}

To characterize the spectral properties of the down-converted photons, we measure the joint spectral intensity~(JSI) of the two photon pairs separately. As shown in Fig.~\ref{FigS5:JSI} (a), the PPLN waveguide is simultaneously pumped by a 775~nm pulsed laser and a tunable telecom CW laser, occuring difference frequency generation~(DFG). The wavelength of the telecom laser is scanned within the range of the CWDM channel corresponding to the signal photon. After filtering out the pump light, the generated idler photon are directed into a spectrometer to measure its spectral information. In Fig.~\ref{FigS5:JSI} (b), the tunable telecom CW laser scan from 1515~nm to 1540~nm, and the spectrometer measure the spectrm from CWDM channel filters centered at 1570~nm. In Fig.~\ref{FigS5:JSI} (c), the tunable telecom CW laser scan from 1500~nm to 1525~nm, and the spectrometer measure the spectrm from CWDM channel filters centered at 1590~nm. From Fig.~\ref{FigS5:JSI} (b) and (c), for each pair of wavelength channels we observe two DFG peaks via the spectrometer, which we attribute to inhomogeneous waveguide structures—a characteristic also manifested in the SHG measurements.

\section{More Experimental details}

\subsection{Experimental details about the generation of multiphoton states}

\begin{figure}[htbp]
\includegraphics[width=0.5\linewidth]{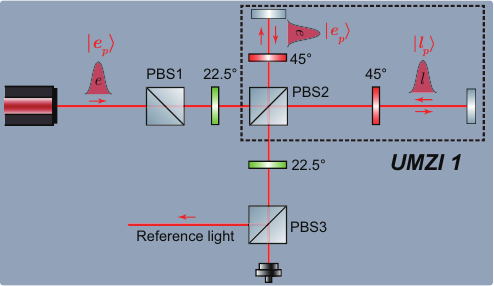}
\caption{Experimental setup for preparing and modulating the pump light  to generate the state  $\ket{\psi_p}=\frac{1}{\sqrt{2}}\left(\ket{e_p}+e^{i\phi_p}\ket{l_p}\right)\otimes\ket{H_p}$.}
\label{FigS6:PulseGeneration}
\end{figure}

\begin{figure}[htbp]
\includegraphics[width=0.5\linewidth]{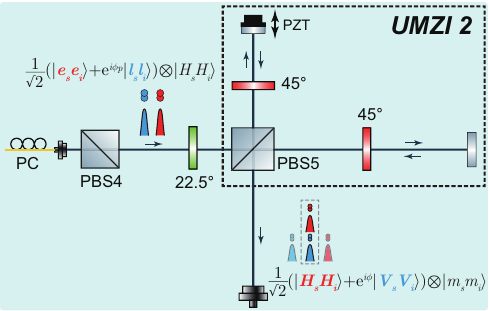}
\caption{Experimental setup for converting time-bin entanglement $\frac{1}{\sqrt{2}} \left( \ket{e_s e_i} + e^{i\phi_p}\ket{l_s l_i} \right) \otimes \ket{V_s V_i}$ to polarization entanglement $\frac{1}{\sqrt{2}} \left( \ket{H_s H_i} + e^{i\phi}\ket{V_s V_i} \right) \otimes \ket{m_s m_i}$. }
\label{FigS7:DOFConversion}
\end{figure}

We describe the generation process of the multiphoton state in detail through step-by-step calculations. First, a pulsed laser is modulated and prepared into the pump light required for state generation through the first unbalanced Mach–Zehnder interferometer~(UMZI 1), as shown in Fig.~\ref{FigS6:PulseGeneration}. A pulse from the laser passes through a polarizing beam splitter~(PBS1), carrying horizontal polarization. A half-wave plate~(HWP) set at 22.5$\degree$ modulates the pulse into a superposition state of horizontal~($\ket{H_p}$) and vertical~($\ket{V_p}$) polarizations. PBS2 direct the $\ket{H_p}$ and $\ket{V_p}$ polarization into the long and short arms of UMZI 1, placing them in time bins $\ket{e_p}$ and $\ket{l_p}$, respectively. Their polarizations are flipped by passing through a quarter-wave plate~(QWP) set at 45$\degree$ twice. These two polarization components are then recombined by PBS2, resulting in a coherent superposition of the pulses with a relative phase $\phi_p$. The modulated pulse undergoes polarization rotation by an HWP set at 22.5$\degree$, after which it is split into two parts by PBS3. The reflected part, with $\ket{V_p}$ polarization, is used as the reference light. The transmitted part, with $\ket{H_p}$ polarization, is coupled into an SMF as the pump light. The pump pulse generation process is described by Eq.~\ref{eq:PulseGeneration}.
\begin{equation}
\begin{aligned}
  \ket{e_p} &\xrightarrow{\mathrm{PBS}1} \ket{e_p} \otimes \ket{H_p} \\
&\xrightarrow{\mathrm{HWP}@22.5\degree} \frac{1}{\sqrt{2}} \left( \ket{e_p} \otimes \ket{H_p} + \ket{e_p} \otimes \ket{V_p} \right) \\
&\xrightarrow[\text{UMZI}~1]{\mathrm{PBS}2 + \mathrm{QWP}@45\degree + \mathrm{PBS}2} \frac{1}{\sqrt{2}}\left( e^{i\phi_p}\ket{l_p} \otimes \ket{V_p} + \ket{e_p} \otimes \ket{H_p}  \right) \\
&\xrightarrow{\mathrm{HWP}@22.5\degree} \frac{1}{2}\left( e^{i\phi_p}\ket{l_p} \otimes \ket{H_p} - e^{i\phi_p}\ket{l_p} \otimes \ket{V_p} + \ket{e_p} \otimes \ket{H_p} + \ket{e_p} \otimes \ket{V_p} \right) \\
&\xrightarrow{\mathrm{PBS}3} \frac{1}{2} \left[ \left( \ket{e_p} + e^{i\phi_p}\ket{l_p} \right) \otimes \ket{H_p} \right]_t + \frac{1}{2} \left[ \left( \ket{e_p} - e^{i\phi_p}\ket{l_p} \otimes \ket{V_p} \right) \right]_r.  \\  
\end{aligned}
\label{eq:PulseGeneration}
\end{equation}
Subsequently, the polarization of the pump light is rotated to $\ket{V_p}$ by a polarization controller~(PC) to drive type-0 SPDC processes, thereby generating time-bin entangled states. Eq.~\ref{eq:StateGeneration} describes this process in detail.
\begin{equation}
\begin{aligned}
& \frac{1}{\sqrt{2}}\left( \ket{e_p} + e^{i\phi_p}\ket{l_p} \right) \otimes \ket{H_p} \\
&\xrightarrow{\mathrm{PC}}  \frac{1}{\sqrt{2}}\left( \ket{e_p} + e^{i\phi_p}\ket{l_p} \right) \otimes \ket{V_p} \\
&\xrightarrow{\mathrm{PPLN}}  \frac{1}{\sqrt{2}}\left( \ket{e_s e_i} + e^{i\phi_p}\ket{l_s l_i} \right) \otimes \ket{V_s V_i}. 
\label{eq:StateGeneration}
\end{aligned}
\end{equation}
To characterize the time-bin entangled state, UMZI 2 is used to convert time-bin entanglement to polarization entanglement, as illustrated in Fig.~\ref{FigS7:DOFConversion}. The generated signal ($s$) and idler ($i$) photon pass through a PC and PBS4, carrying $\ket{V_s V_i}$ polarization. An HWP set at 22.5$\degree$ rotates them into an equally superposed polarization state. PBS5 splits each photon into $\ket{H}$ and $\ket{V}$ components, directing them into the long and short arms of UMZI 2, respectively. The two components are placed into new time bins with their polarization flipped in UMZI 2. The path-length difference of the two UMZIs is precisely matched, ensuring temporal overlap between the relevant time bins and enabling coherent interference. Consequently, there are four possible time bins after UMZI 2: $\ket{ee}$, $\ket{el}$, $\ket{le}$, $\ket{ll}$. Among them, $\ket{el}$ and $\ket{le}$ are redefined as time bin $\ket{m}$ as they have indistinguishable overlap. There are also phases $\phi_s$ and $\phi_i$ arising from the optical path-length differences experienced by the signal and idler photons in UMZI 2. Using the pulse train of the pump laser as a timing reference, $\ket{m}$ is post-selected as the time label for the polarization-entangled state. This time-bin post-selection has a 50\% success probability for each photon, corresponding to 3~dB loss per channel. After the degree-of-freedom~(DOF) conversion, the polarization-entangled state contains a new relative phase $\phi = \phi_s + \phi_i - \phi_p$. Eq.~\ref{eq:DOFConversion} describes the DOF conversion process in detail.

\begin{equation}
\begin{aligned}
& \frac{1}{\sqrt{2}} \left( \ket{e_s e_i} + e^{i\phi_p}\ket{l_s l_i} \right) \otimes \ket{V_s V_i} \\
& \xrightarrow{\mathrm{PC}+\mathrm{PBS}4} \frac{1}{\sqrt{2}} \left( \ket{e_s e_i} + e^{i\phi_p}\ket{l_s l_i} \right) \otimes \ket{H_s H_i} \\
& \xrightarrow{\mathrm{HWP}@22.5\degree} \frac{1}{2\sqrt{2}} \left( \ket{e_s e_i} + e^{i\phi_p}\ket{l_s l_i} \right) \otimes \left( \ket{H_s H_i} + \ket{H_s V_i} + \ket{V_s H_i} + \ket{V_s V_i} \right) \\
& \qquad \qquad = \frac{1}{2\sqrt{2}} \left( \ket{e_s e_i} + e^{i\phi_p}\ket{l_s l_i} \right) \otimes \ket{H_s H_i} + \frac{1}{2\sqrt{2}} \left( \ket{e_s e_i} + e^{i\phi_p}\ket{l_s l_i} \right) \otimes \ket{H_s V_i} \\
& \qquad \qquad \quad + \frac{1}{2\sqrt{2}} \left( \ket{e_s e_i} + e^{i\phi_p}\ket{l_s l_i} \right) \otimes \ket{V_s H_i} + \frac{1}{2\sqrt{2}} \left( \ket{e_s e_i} + e^{i\phi_p}\ket{l_s l_i} \right) \otimes \ket{V_s V_i} \\
& \xrightarrow[\text{UMZI}~2]{\mathrm{PBS}5 + \mathrm{QWP}@45\degree + \mathrm{PBS}5} \frac{1}{2\sqrt{2}} \left( e^{i\phi_s}e^{i\phi_i}\ket{el_s el_i} + e^{i\phi_p}e^{i\phi_s}e^{i\phi_i}\ket{ll_s ll_i} \right) \otimes \ket{V_s V_i} \\
& \qquad \qquad \qquad \qquad \qquad + \frac{1}{2\sqrt{2}} \left( e^{i\phi_s}\ket{el_s ee_i} + e^{i\phi_p}e^{i\phi_s}\ket{ll_s le_i} \right) \otimes \ket{V_s H_i} \\ 
& \qquad \qquad \qquad \qquad \qquad + \frac{1}{2\sqrt{2}} \left( e^{i\phi_i}\ket{ee_s el_i} + e^{i\phi_p}e^{i\phi_i}\ket{le_s ll_i} \right) \otimes \ket{H_s V_i} \\
& \qquad \qquad \qquad \qquad \qquad + \frac{1}{2\sqrt{2}}\left( \ket{ee_s ee_i} + e^{i\phi_p}\ket{le_s le_i} \right) \otimes \ket{H_s H_i} \\
& \xrightarrow{\text{Postselection}} \frac{1}{\sqrt{2}} e^{i\phi_p}\ket{le_s le_i} \otimes \ket{H_s H_i} + \frac{1}{\sqrt{2}} e^{i\phi_s}e^{i\phi_i}\ket{el_s el_i} \otimes \ket{V_s V_i}  \\
& \xrightarrow{\ket{le_s le_i}, \ket{el_s el_i} \rightarrow \ket{m_s m_i}} \frac{1}{\sqrt{2}} \left( \ket{H_s H_i} + e^{i\phi}\ket{V_s V_i} \right) \otimes \ket{m_s m_i}. 
\end{aligned}
\label{eq:DOFConversion}
\end{equation}

\subsection{System Losses}

The per-channel photon loss in the four-photon state measurement totals 15.5~dB. We summarize the losses of each component of the system in Table.~\ref{Tb:Loss}. The major loss is a fiber-to-chip coupling loss of ~4 dB, which severely affects multi-photon source performance. By taper designs and optimized fabrication processes could reduce the coupling loss less than 0.5~dB. Another non-negligible loss is the DOF conversion loss and this can be optimized via deterministic time-bin analyzers using electro-optic modulation.
Besides, the loss of the UMZI insertion and polarization analyzer could also be further optimized. After the improvements mentioned above, the count rate of multi-photon quantum light sources will be substantially enhanced without compromising fidelity.

\begin{table}[htbp]
\renewcommand\arraystretch{1.5}
\centering
\begin{threeparttable}
\caption{Loss analysis for the entangled state measurement system.}
\label{Tb:Loss}
    \begin{tabular}{p{6cm}<{\centering}p{1.5cm}}
    \hline\hline
    Fiber-to-chip coupling loss
    & 4~dB \\
    \hline
    UMZI insertion loss
    & 2.7~dB \\
    \hline
    DOF conversion loss
    & 3~dB \\
    \hline
    CWDM insertion loss
    & 1.5~dB \\
    \hline
    Polarization analyzer insertion loss
    & 2.5~dB \\ 
    \hline
    Transmission loss to SNSPD
    & 0.8~dB \\
    \hline
    Photon detection loss
    & 1~dB \\
    \hline
    \hline
    \end{tabular} 
    \end{threeparttable}
\end{table}

\section{Four-photon interference measurement}
In our experiment, four photon quantum state if formed by two simultaneous Bell states, with its ideal expression given by $\ket{\Phi_4}=\ket{\Phi_2}^{\otimes2}$. During the measurment of four-photon polarization correlations, the angles of HWPs in the signal photons are fixed at $\theta_{s_1}=\theta_{s_2}=\theta_{s}$, while the angles of HWPs in the idler channels $i_1$ and $i_2$ are rotated synchronously from $0^{\circ}$ to $180^{\circ}$, i.e., $\theta_{i_1}=\theta_{i_2}=\theta_{i}$ vary from $0^{\circ}$ to $180^{\circ}$. 

Considering the Jones matrix of the HWP
\begin{equation}
U_\text{\tiny HWP}(\theta)=\left(\begin{array}{cc}
\cos 2 \theta & \sin 2 \theta \\
\sin 2 \theta & -\cos 2 \theta
\end{array}\right),
\end{equation}
the projection for signal photons is $\ket{\psi_{s1}}=\ket{\psi_{s2}}=U_\text{\tiny HWP}(\theta_s)\ket{H}$ and the projection for idler photons is $\ket{\psi_{i1}}=\ket{\psi_{i2}}=U_\text{\tiny HWP}(\theta_i)\ket{H}$, so the composite projection operator is given by $\ket{\Psi}=\ket{\psi_{s1}}\ket{\psi_{i1}}\ket{\psi_{s2}}\ket{\psi_{i2}}$. This leads to an expected four-photon interference proportional to $P=|\bra{\Psi}\Phi_4\rangle|^2=\frac{1}{4}\mathrm{cos}^4\left( 2\theta_i-2\theta_s \right)$.

\section{Comparison of integrated four photon source}

We summarize previously reported on-chip generation of two simultaneous Bell states based on other material platforms. In comparison, our results enhance the four-fold coincidence rate by approximately a factor of three.

\begin{table}[htbp]
\renewcommand\arraystretch{1.5}
\centering
\begin{threeparttable}
\caption{A summary of four photon quantum sources (two simultaneous Bell states) on chip.}
\label{Tb:comparasion}
    \begin{tabular}{p{2.5cm}<{\centering}p{6cm}<{\centering}p{4cm}<{\centering}p{2cm}<{\centering}}
    \hline\hline
    ~  
    & Platform 
    & Four-fold coincidence rate 
    & Fidelity \\ 
    \hline
    This work
    & LNOI 
    & 1~Hz
    & 0.74 \\
    \hline
    Reimer~\emph{et al.}~\cite{reimer2016Sa}
    & High refractive index silica glass (Hydex)
    & 0.17~Hz
    & 0.64 \\
    \hline
    Zhang~\emph{et al.}~\cite{zhang2019LSAa}
    & Si 
    & 0.34~Hz
    & 0.78  \\
    \hline
    Lee~\emph{et al.}~\cite{Lee2024APL}
    & Si  
    & 0.2~Hz
    & 0.874  \\
    \hline
    \hline
    \end{tabular} 
    \end{threeparttable}
\end{table}

\twocolumngrid

\bibliography{Ref}

\end{document}